\documentclass[preprint,12pt]{elsarticle}

\usepackage{graphicx}

\usepackage{amssymb}
\usepackage{amsmath}

\journal{Nuclear Physics B}

\begin{document}

\begin{frontmatter}



\title{Scale dependence of top-quark cross section at $e^+ e^-$ colliders  near production threshold at NNNLO}


\author{Yuichiro Kiyo} 

\affiliation{organization={Department of Physics, Juntendo University},
            city={Inzai},
            state={Chiba},
            postcode={270-1695}, 
            country={Japan}}

\begin{abstract}
The top-quark threshold cross section at \(e^+e^-\) colliders near
production threshold is investigated. We study the scale dependence of the
cross section for $\sigma(e^+e^- \to t\bar{t}X)$ near $\sqrt{s}\simeq 2m_t$ 
and discuss the theoretical accuracy of the NNNLO prediction.
We report that a threshold scan at an $e^+ e^-$ collider would allow a 
determination of the top-quark mass with an accuracy of order $30\, \mathrm{MeV}$.
\end{abstract}




\end{frontmatter}



\section{Introduction}

The top quark is the heaviest fundamental particle in the Standard Model (SM), 
and a precise determination of its mass and couplings is crucial for precision
tests of electroweak and QCD dynamics and for assessing the stability of 
the SM vacuum. 

The top-quark production cross section at $e^+  e^-$ colliders near threshold 
is a key observable for precision studies of top-quark properties. 
To match the anticipated experimental accuracy,  it is mandatory to provide 
theoretical predictions for  the top-quark cross section with commensurate precision.
More than 20 years ago, the NNLO cross section was computed by several groups 
and was found to suffer from rather large theoretical uncertainties \cite{Hoang:2000yr}.  
It was understood that part of this uncertainty originates from the renormalon ambiguity
 in the pole-mass definition, which can be avoided by using suitable short-distance
  mass schemes.
  
 However, even after this improvement, the theoretical uncertainty at NNLO did not
  become sufficiently small. Subsequently, attempts were made to stabilize the 
  threshold cross section by resumming logarithms of $v$ (speed of top quark) at NNLL
  accuracy  \cite{Hoang:2001mm,Pineda:2006ri}. Nonetheless, 
  the inclusion of the NNNLO non-logarithmic terms, which are not contained 
  in the NNLL approximation, is required to ensure that the theoretical prediction is
sufficiently accurate. Therefore we have chosen to pursue a direct computation of
 the QCD corrections at NNNLO. As a result of this effort, the top-quark production
 cross section is now known up to NNNLO, as reported in Refs.~\cite{Beneke:2015kwa,Beneke:2016kkb}. 
 
 In this article,  we study the scale dependence of the threshold cross section based
  on the detailed calculations presented in Refs.~\cite{Beneke:2013jia,Beneke:2024sfa}, 
  and update the theoretical uncertainties.

\section{Threshold production cross section}

Using the optical theorem, the top-quark production cross section $\sigma_{t\bar{t}}$ for
$e^+e^- \to \gamma^\ast/Z^\ast \to t\bar{t}$ can be written in terms of the
imaginary part of the current--current correlation functions
$\Pi_{\mu\nu}^{(X)}(q)
=
i \int d^4x \, e^{iqx}
\langle \mathrm{T}\ j_\mu^{(X)}(x) j_\nu^{(X)}(0) \rangle$,
with $X=v,a$ denoting the vector and axial-vector currents, respectively.
For instance, the photon-mediated contribution to the normalized 
cross section (the $R$-ratio) is given by
\begin{align}
R_{t\bar{t}}
&=\frac{\sigma_{t\bar{t}}}{\sigma_0}
=
12\pi e_t^2
\, \mathrm{Im}\, \Pi^{(v)}(q),
\end{align}
where $s=q^2$ is the squared center-of-mass energy.  
Near the production threshold, $\sqrt{s} \sim 2m_t$, the correlation functions
can be computed using NRQCD, and $\Pi^{(v)}(q)$ can be expressed in terms of the
nonrelativistic Green function $G(E)$ as
\begin{align}
\Pi^{(v)}(q)
&=
\frac{2N_c}{s}
\bigg[
c_v \left( c_v - \frac{d_v}{3}\frac{E}{m_t} \right)
G(E)\bigg]
\label{eq:Pi_v}
\end{align}
where $E = \sqrt{s} - 2m_t$ is the energy of the top-quark pair, 
with $m_t$ being the top-quark mass. The coefficients $c_v, d_v$ are the matching
 coefficients of the vector current. Since the matching between NRQCD and QCD is performed for  on-shell top quarks, 
 $\Pi^{(v)}(q)$ in \eqref{eq:Pi_v} describes the so-called resonant production
 of a top-quark pair. Accordingly, we identify the corresponding cross section as
  $R_{t\bar{t}}\equiv R_{\rm res}$, which we refer to as the resonant cross section.

Owing to the large top-quark mass,  the top quark has a large decay width $\Gamma_t$, which gives rise to a smearing of the resonance peaks of top-quark bound states in the top-quark cross section. The top-quark width effect can be taken into account 
by the well-known prescription: replacing the energy $E$ in \eqref{eq:Pi_v}  
with the complex energy $\mathcal{E}=E+i\Gamma_t$.
Here, it is important to realize that the large imaginary part $i\Gamma_t$ of the complex energy $\mathcal{E}$ allows the cross section to probe far off-shell top-quark momentum regions. 
This fact is manifested by the appearance of ultraviolet divergences proportional to 
$i\Gamma_t$, which signal contributions from non-resonant momentum regions. 

For a finite value of $i\Gamma_t$,  these non-resonant 
contributions must be included in order to obtain a well-defined cross section. Otherwise, there would be no counter-term that  cancels the divergences associated with $i\Gamma_t$, and the theoretical description would break down.
Therefore, we add the contributions from non-resonant regions to 
$R_{\rm res}$ and define the physical cross section as
\begin{align}
R
&=R_{\rm res}(\mu,\mu_w)+R_{\rm non-res}(\mu', \mu_w),
\end{align}
where $\mu$ and $\mu'$ are the renormalization scales of the QCD 
coupling constant $\alpha_s$ in the respective cross sections, 
and $\mu_w$ is the  factorization scale for the resonant and 
non-resonant momentum regions. We refer to $\mu_w$ as the finite-width scale.
The dependences on $\mu$ and $\mu'$  cancel order by order 
in the perturbative expansion in $\alpha_s$ within $R_{\rm res}$ 
and $R_{\rm non-res}$, respectively. The finite-width scale dependence
 cancels only when resonant and non-resonant cross sections are consistently included.
  At present, the non-resonant corrections  are known up to NNLO \cite{Beneke:2017rdn}.
We therefore do not consider them further in this work and focus on the 
scale dependence of the resonant cross section in the next section.

\section{Scale dependence of cross section}

\begin{figure}[htbp]
\centering
\includegraphics[bb=0 0 1080 720, width=0.45\linewidth]{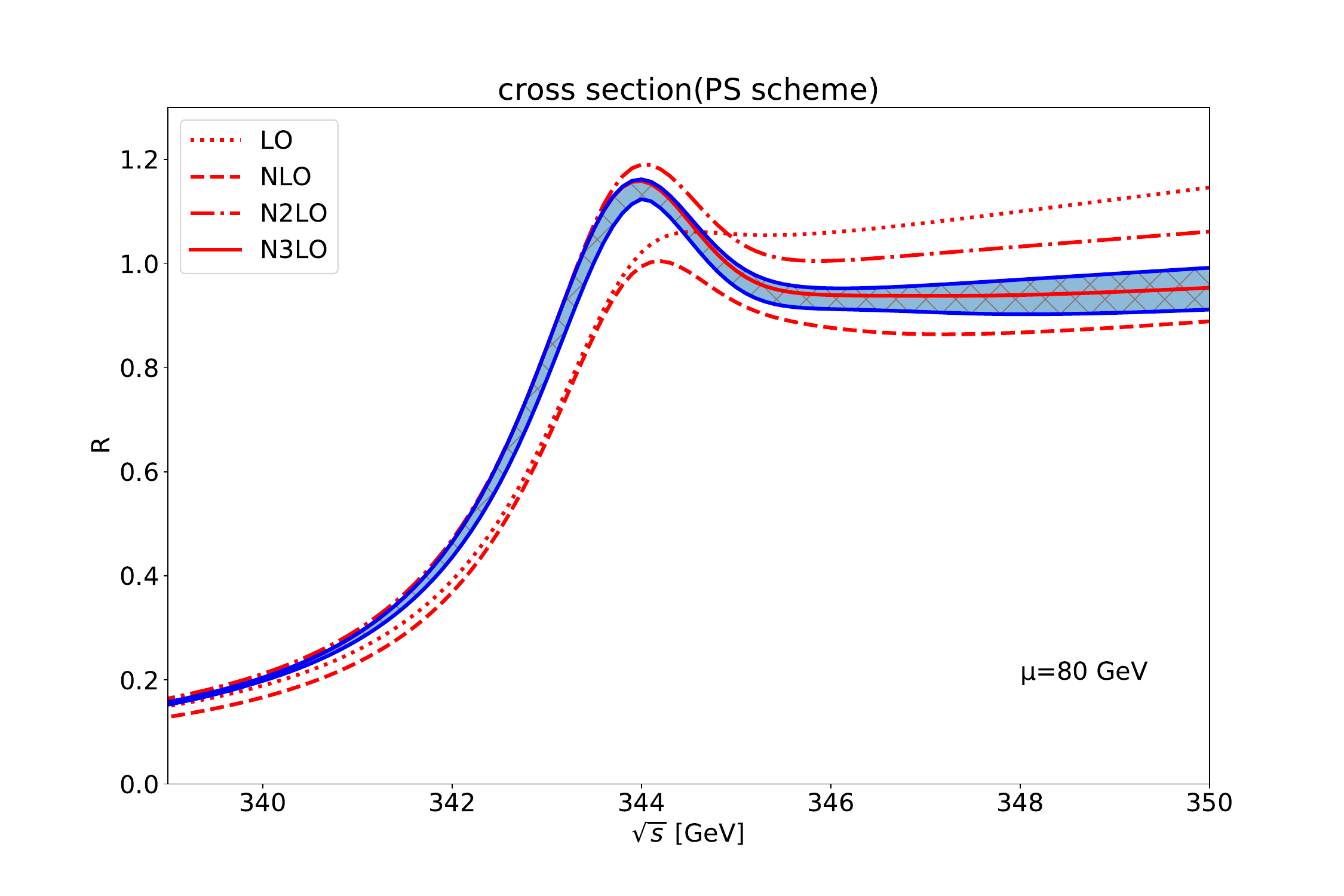}
\includegraphics[bb=0 0 1080 720, width=0.45\linewidth]{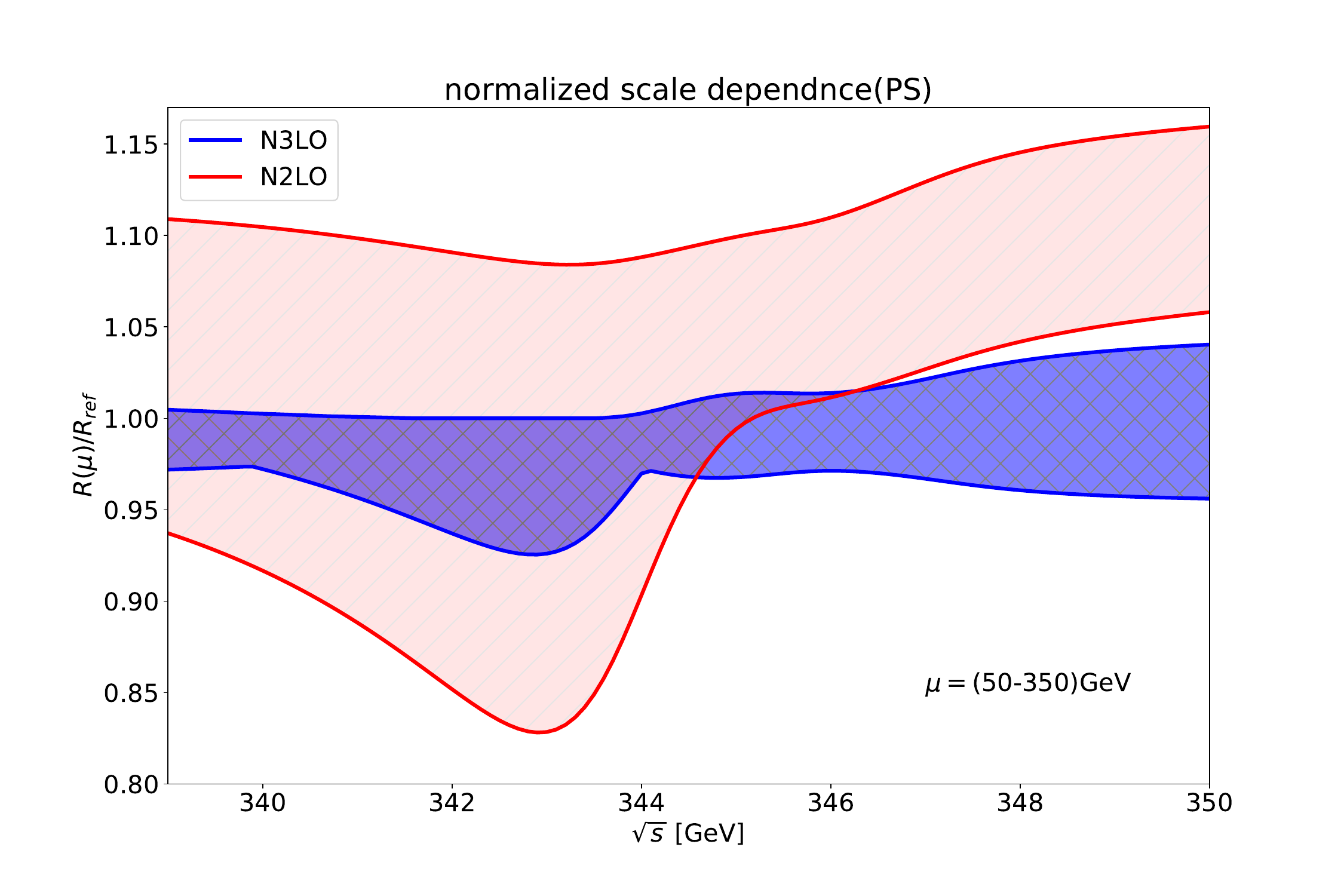}
\caption{Top-quark production cross section in QCD near threshold with 
$\mu=80~{\rm GeV}$ (left panel), and the relative renormalization-scale dependence (right panel). Here $R$ denotes the resonant contribution only.}\label{fig1}
\end{figure}
In Fig.~\ref{fig1}, we show the resonant cross section including QCD corrections 
\footnote{In addition to QCD, QED and electroweak corrections can be 
included using the \texttt{QQbar\_threshold} library developed in Ref.~\cite{Beneke:2016kkb}.}
order by order (left panel), 
evaluated at $\mu=80~{\rm GeV}$ and $\mu_w=350~{\rm GeV}$, 
and the relative scale dependence obtained by varying $\mu$ in the range $\mu\in [50, 350]~{\rm GeV}$ (right panel) with the finite-width scale fixed to 
$\mu_w=350~{\rm GeV}$. The blue band represents the scale variation of the NNNLO
 cross section.  We adopted $m_{t}^{\rm PS}(\mu_f=20~{\rm GeV})=171.5~{\rm GeV}, 
\Gamma_t=1.36~{\rm GeV}$, and $\alpha_s(m_Z)=0.1180$. 
  These results demonstrate perturbative convergence 
of the top-quark resonant cross section with respect to QCD corrections, 
 and the renormalization-scale uncertainty is reduced from about 20\% at 
 NNLO  to about  5\%  at NNNLO.

\begin{figure}[t]
\centering
\includegraphics[bb=0 0 360 259, width=0.3\linewidth]{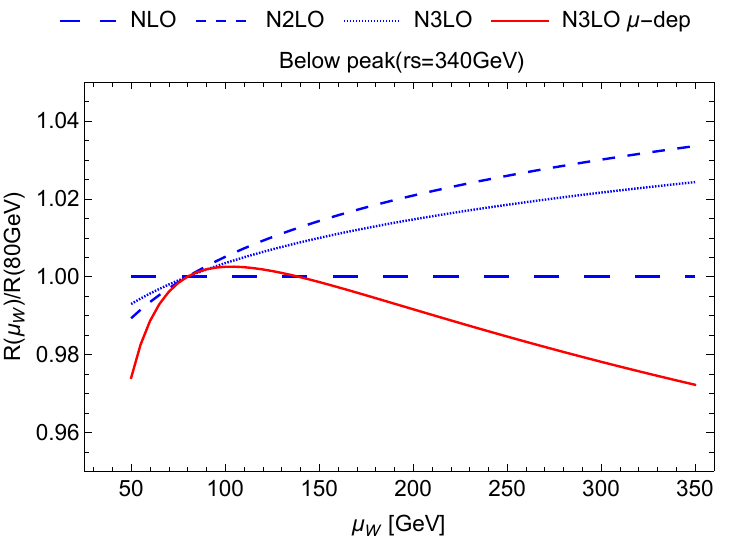}
\includegraphics[bb=0 0 360 259, width=0.3\linewidth]{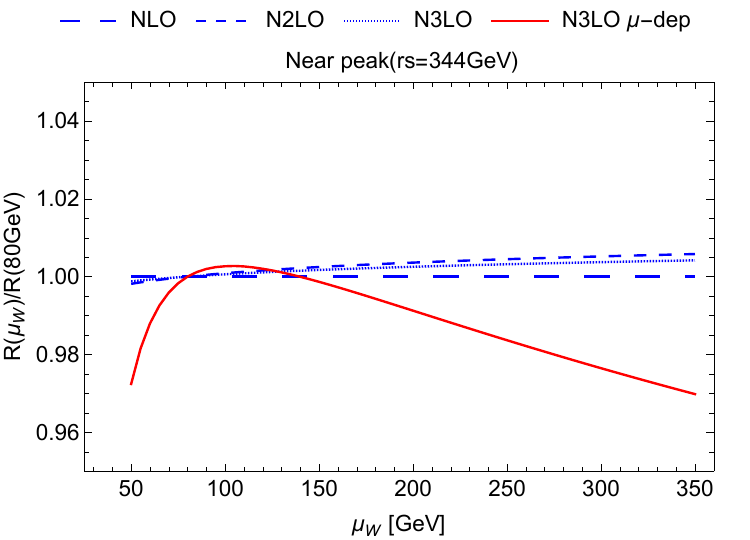}
\includegraphics[bb=0 0 360 259, width=0.3\linewidth]{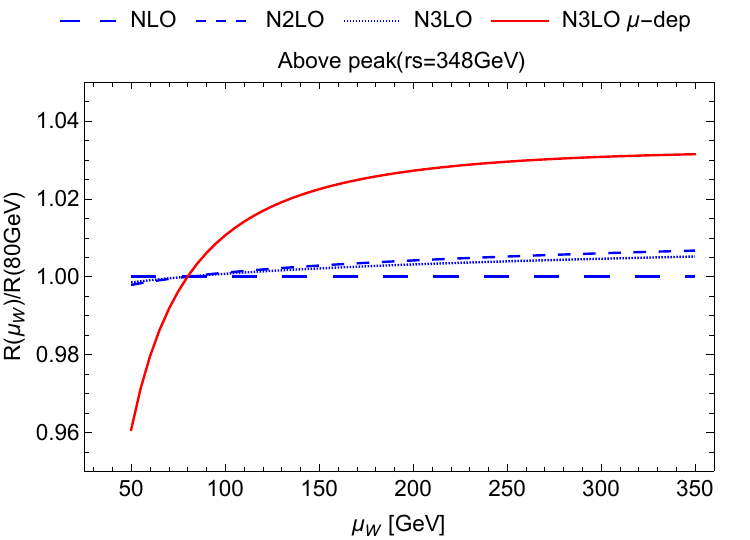}
\caption{Finite-width scale dependence of the resonant top-quark cross section 
at NLO, NNLO, and NNNLO for center-of-mass energies
$\sqrt{s}=340 , 344$ , and $348 ~{\rm GeV}$,  shown in 
the left, middle, and right panels, respectively.
}\label{fig2}
\end{figure}
For the resonant QCD cross section, the $\mu_w$-dependence starts to
 appears at NNLO and cancels only when non-resonant cross sections 
 at the same order are included. In Fig.~\ref{fig2}, we show the 
$\mu_w$-dependence of the normalized resonant cross section, 
 $R(\mu=80~{\rm GeV},\mu_w)/R(\mu=\mu_w=80~{\rm GeV})$ 
 for $\sqrt{s}= 340, 344$, and $348~{\rm GeV}$ 
  in the range $\mu_w \in [50, 350]~{\rm GeV}$. 
  These center-of-mass energies correspond to
 representative points below the peak, at the peak, and above the peak,
 respectively, as can be seen from the resonant cross section shown 
 in Fig.~\ref{fig1}. The $\mu_w$-dependence 
  is rather  small,  at the level of  1\% near the peak  of the cross section at 
  $\sqrt{s}=344~{\rm GeV}$ (middle panel).
   For comparison, the $\mu$-dependence of the NNNLO 
   cross section is shown by the solid red line in each panel. 
   Compared to the $\mu$-dependence , the $\mu_w$-dependence
   is significantly weaker around the peak and at higher center-of-mass
   energies, however, the $\mu_w$-dependence becomes relatively more 
   pronounced at $\sqrt{s}=344~{\rm GeV}$. 
   This is because  the resonant cross section decreases rapidly 
   in this region, while the non-resonant cross section is largely independent 
   of $\sqrt{s}$ and thus gives a relatively larger contribution.

\begin{figure}[t]
\centering
\includegraphics[bb=0 0 1218 558, width=0.8\linewidth]{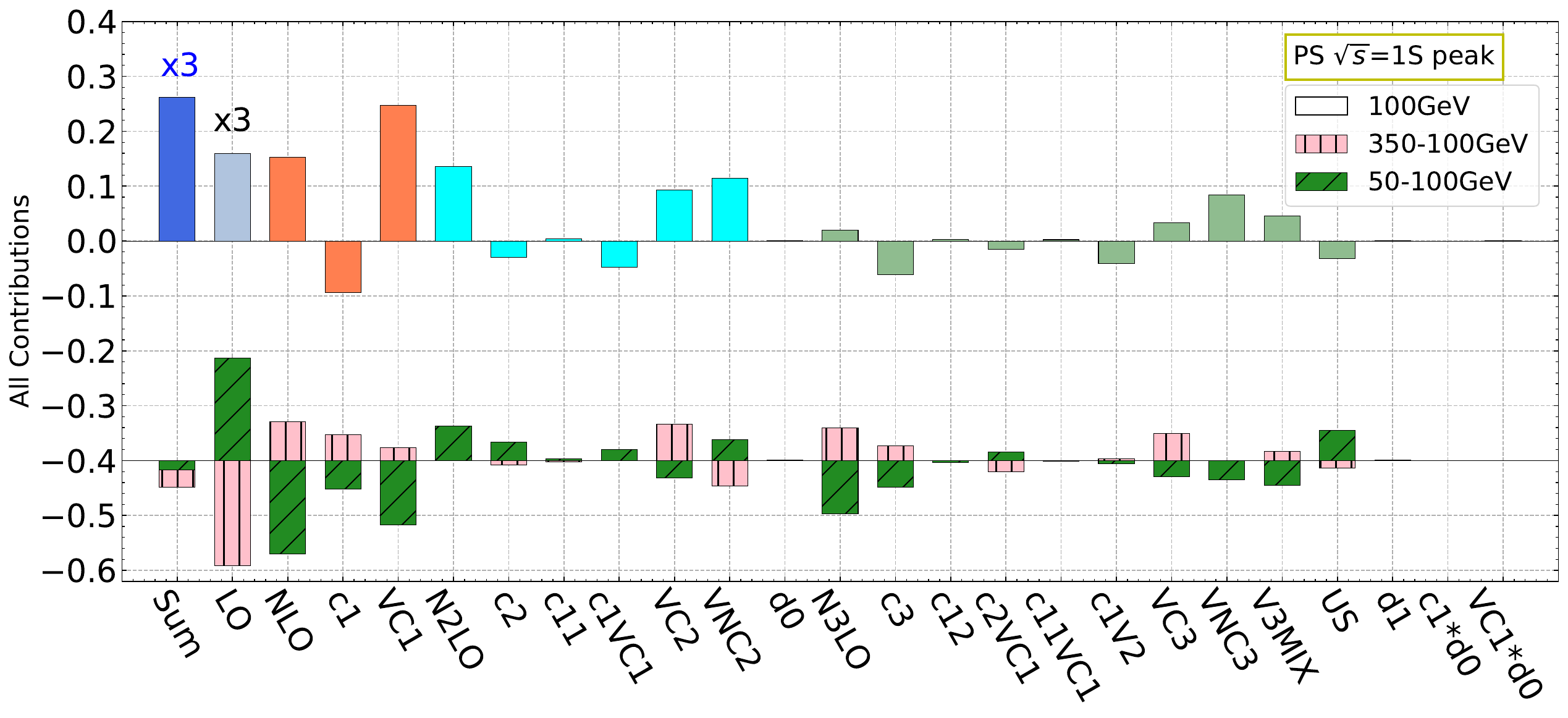}
\caption{Size of perturbative QCD corrections to the $R$-ratio 
at $\sqrt{s}=344~{\rm GeV}$ with $\mu=100~{\rm GeV}$ 
(upper panel), and their scale dependence for $\mu \in [50, 350]~{\rm GeV}$ 
(lower panel).
}\label{fig3}
\end{figure}
In Fig.~\ref{fig3}, we show the size of the individual QCD corrections to the
 $R$-ratio at $\sqrt{s}=344~{\rm GeV}$ (1S peak) and their $\mu$ dependence.
 The upper panel displays the relative size of the 
corrections at $\mu=100~{\rm GeV}$  from individual potentials and matching coefficients, 
which are listed on the horizontal axis. The contributions are grouped 
according to the perturbative order:
NLO contributions are shown in orange, NNLO contributions in cyan, 
and NNNLO contributions in light green. The lower panel displays the scale 
variation for $\mu\in [50, 350]~{\rm GeV}$. The scale variation in the range 
$\mu \in [50, 100]~{\rm GeV}$ is indicated by the green bars, while that in the
range $\mu \in [100, 350]~{\rm GeV}$ is shown by the pink bars.

The label ``Sum" denotes the
 sum of all contributions up to NNNLO.  At LO, the correction consists solely
  of the Coulomb potential. The ``Sum" and LO bars are rescaled by a factor $1/3$.
  
At NLO, the corrections contain two contributions,  $\texttt{c1}$ and $\texttt{VC1}$, 
and their sum is $\texttt{NLO}$.
At NNLO, six contributions are included. 
At NNNLO, the corrections consist of twelve individual contributions, 
$\texttt{c3}, \cdots  ,\texttt{VC1*d0}$, and  their sum is $\texttt{N3LO}$.
From the figure, the sum of the NNNLO contributions
is small with $\texttt{N3LO}=0.02$ at $\mu=100~{\rm GeV}$, 
but its scale dependence is rather strong, 
amounting to $+0.06$ for 
$\mu\in [100, 350]~{\rm GeV}$ (pink bar)
 and $-0.10$ for $\mu \in [50,100]~{\rm GeV}$ (green bar).

In this way, the contributions from individual perturbative orders 
exhibit a strong scale dependence. However, in their sum, 
a substantial cancellation of the scale dependence occurs. 
As a result, the total contribution to $R$ is $0.80$ at $\mu=100~{\rm GeV}$, with 
variation of $-0.04$ for $\mu \in [100, 350]~{\rm GeV}$ 
and $-0.02$ for $\mu \in [50,100]~{\rm GeV}$.

\begin{table}[!htbp]
\newcommand{\m}{\hphantom{$-$}}
\newcommand{\cc}[1]{\multicolumn{1}{c}{#1}}
\renewcommand{\tabcolsep}{0.6pc} 
\renewcommand{\arraystretch}{0.95} 
\caption{Shift of the peak position relative to the previous 
order and its $\pm$ scale variation in the three short distance 
mass schemes. All the numbers are  in units of MeV.
At NNNLO, the peak is located at 
$\sqrt{s}=$ 343.972 (PS scheme),  343.985 ($\overline{\rm MS}$ scheme),
 343.972 (PS scheme, $\mu_f=50~{\rm GeV}$) 
GeV.
}
\label{tab:peak}
\begin{center}
\begin{tabular}{@{}clllll}
\hline  \vspace{-4mm}\\  
\mbox{Order} &  \hspace{-5mm}   
     & \mbox{PS scheme} 
     & \mbox{$\overline{\rm MS}$ scheme}
     & \hspace{-7mm} \mbox{PS scheme}, $\mu_f=50~{\rm GeV}$
\\
\hline   \\[-2mm]
NLO  & \hspace{-5mm} 
     & $-367 / \pm 145$
     & $+1008 / \pm 482\;\;$
     & $+154 / \pm 44$
 \vspace{2mm} \\
NNLO  & \hspace{-5mm} 
     & $-149 / \pm 107$
     & $+46 / \pm 103$
     & $+2 / \pm 16$
 \vspace{2mm} \\
NNNLO  & \hspace{-5mm} 
     & $-65 / \pm 61$
     & $-81 / \pm 26$
     & $-9 / \pm 13$
 \vspace{2mm} \\
\hline
\vspace{-10mm}
\end{tabular}
\end{center}
\end{table} 
Finally, we discuss the QCD corrections to the position of the cross-section
peak, which approximately corresponds to the $1S$ resonance energy.
In Table~\ref{tab:peak}, we show the shift of the 
 peak position with respect to the previous perturbative order, 
 together with the corresponding $\pm$ scale variation,
in three different short-distance mass schemes. 
 For example, in the PS scheme, the NNNLO correction amounts to
  $-65~{\rm MeV}$, with an associated scale uncertainty 
  of $\pm 61~{\rm MeV}$.
  
  It is expected that higher-order corrections beyond NNNLO are smaller than
this value. Since the peak position scales approximately with $2m_t$,
it is reasonable to take about half of this scale dependence as an estimate
of the quark-mass uncertainty when using the NNNLO cross section.
Consequently, a threshold scan at an $e^+e^-$ collider would allow the top-quark
mass to be determined with an accuracy of about $30~{\rm MeV}$, given the
current precision of the cross-section calculation.

\section{Conclusion}
 We have computed the top-quark threshold production cross section up to
NNNLO accuracy. In this work, we have focused on QCD effects; however,
phenomenological analyses can be performed using codes that also include
the currently known QED and electroweak corrections. The theoretical
uncertainty of the NNNLO QCD cross section is at the level of about 5\%.
If the top-quark mass is extracted from the position of the cross-section
peak, a precision of $\delta m_t \sim 30\,\mathrm{MeV}$ is expected to be
achievable.


\section*{Acknowledgements}
The  author would like to thank the organizers of the TOP2025 workshop.  
The author is grateful to M. Beneke for a long-term collaboration.
This work was supported in part by the JSPS KAKENHI Grant 
Numbers JP22K03602 and JP25K07325.


\end{document}